\documentclass[aps,prl,preprint,superscriptaddress]{revtex4-1}
\usepackage{amsmath,amsfonts,amsthm,amssymb}
\usepackage{graphicx}
\usepackage{color, soul}
\usepackage[colorlinks=true,linkcolor=black,citecolor=blue,urlcolor=blue]{hyperref}
\begin{document}
\title{Soft nanobubbles can deform rigid glasses}
\author{Shuai Ren}
\thanks{These two authors contributed equally.}
\affiliation{School of Mechanical Engineering and Automation, Beihang University, 37 Xueyuan Rd., Haidian District, Beijing 100191, China.}
\author{Christian Pedersen}
\thanks{These two authors contributed equally.}
\affiliation{Mechanics Division, Department of Mathematics, University of Oslo, 0316 Oslo, Norway.}
\author{Andreas Carlson}
\affiliation{Mechanics Division, Department of Mathematics, University of Oslo, 0316 Oslo, Norway.}
\author{Thomas Salez}
\email{thomas.salez@u-bordeaux.fr}
\affiliation{Univ. Bordeaux, CNRS, LOMA, UMR 5798, F-33405 Talence, France.}
\affiliation{Global Station for Soft Matter, Global Institution for Collaborative Research and Education, Hokkaido University, Sapporo, Hokkaido 060-0808, Japan.}
\author{Yuliang Wang}
\email{wangyuliang@buaa.edu.cn}
\affiliation{School of Mechanical Engineering and Automation, Beihang University, 37 Xueyuan Rd., Haidian District, Beijing 100191, China.}
\affiliation{Beijing Advanced Innovation Center for Biomedical Engineering, Beihang University, 37 Xueyuan Rd., Haidian District, Beijing 100191, China.}
\date{\today}

\begin{abstract}
Confined glasses and their anomalous interfacial rheology raise important questions in fundamental research and numerous practical applications. In this Letter, we study the influence of interfacial air nanobubbles on the free surface of ultrathin high-molecular-weight glassy polystyrene films immersed in water, in ambient conditions. In particular, we reveal the counterintuitive fact that a soft nanobubble is able to deform the surface of a rigid glass, forming a nanocrater with a depth that increases with time. By combining \emph{in situ} atomic-force-microscopy measurements and a modified lubrication model for the liquid-like layer at the free surface of the glass, we demonstrate that the capillary pressure in the nanobubble together with the liquid-like layer at the free surface of the glass determine the spatiotemporal growth of the nanocraters. Finally, from the excellent agreement between the experimental profiles and the numerical solutions of the governing glassy thin-film equation, we are able to precisely extract the surface mobility of the glass. In addition to revealing and quantifying how surface nanobubbles deform immersed glasses, until the latter eventually dewet from their substrates, our work provides a novel, precise, and simple measurement of the surface nanorheology of glasses.
\end{abstract}
\maketitle

The glass transition has been being a major enigma in solid-state physics~\cite{Anderson1995} for almost a century, leading to an important literature for the bulk case~\cite{Berthier2011}. Besides an hypothetical underlying phase transition, the tremendous dynamical slowing down of glass-forming supercooled liquids has been attributed to molecular caging, and the associated requirement for cooperative relaxation~\cite{Adam1965} in a region of a certain cooperative size~\cite{Donth1996}.

The quest for the latter observable, and its possible divergence, led to an alternative strategy: the study of glasses in confinement~\cite{Jackson1990,Forrest2001,Ediger2014}. In the particular case of thin polymer films, anomalies have been reported, such as reductions of the apparent glass-transition temperature $T_{\textrm{g}}$ at small film thicknesses~\cite{Keddie1994,Forrest1996}, where the presence of free surfaces played an important role~\cite{Baumchen2012}. Furthermore, space-dependent $T_{\textrm{g}}$ values were inferred from local measurements~\cite{Ellison2003}. Besides, the free surface of a polymer glass was discovered to be much more mobile than the bulk, which was attributed to the existence of a nanometric liquid-like superficial layer capable to flow under external constraints~\cite{Fakhraai2008,Ilton2009,Yang2010,Chai2014,Kim2018,Ogieglo2018}, or equivalently for small enough molecules to undergo surface diffusion~\cite{Zhu2011,Zhang2016,Tanis2019} as in crystals~\cite{Mullins1957}, which could even lead to striking engulfment phenomena~\cite{LeThe2009}. The previous Stokes-Einstein-like equivalence between surface flow and surface diffusion in the mobile layer was shown to be eventually broken for long-enough surface polymer chains due to their anchoring into the bulk matrix~\cite{Chai2019}, and ultimately the commensurability of their typical size with the sample thickness itself~\cite{de2000glass,Milner2010}. Finally, among other interesting properties, spatial heterogeneities were associated with the dynamics of thin glassy polymer films~\cite{Siretanu2015}. To rationalize these observations, various numerical approaches~\cite{Varnik2002,Baschnagel2005} and theoretical models~\cite{Ngai1998,Long2001,Lipson2009,Mirigian2014,Salez2015,Arutkin2016} have been proposed, but a unifying picture is still at large.

In this Letter, we study the influence of air nanobubbles~\cite{Wang2010,Lohse2015} spontaneously-created at the free surface of ultrathin high-molecular-weight glassy polystyrene (PS) films when immersed in water, and in ambient conditions. In contrast to the bubble-inflation technique used for freestanding viscoelastic membranes~\cite{OConnell2005}, there is here no need for an externally-driven inflation, and the glassy films are supported onto rigid silicon wafers and thus much less compliant. The nanobubbles are gaseous air domains with nanometric height and width. As a consequence of these small sizes, and from the Young-Laplace equation, the pressure inside the bubble can reach up to $\sim10$~bar, which -- despite being much smaller than the yield stress of the bulk glass -- can lead to an external driving force for the flow of the liquid-like layer at the free surface of the glass. Consequently, a nanoscopic crater is formed underneath the bubble, and grows in size with time, as observed using an atomic-force microscope (AFM). The latter observations are discussed in the context of a modified lubrication model for the capillary-driven flow of the liquid-like layer at the free surface of the glassy film, under an external driving force. The excellent agreement between the experimental AFM profiles and the numerical solutions of the axisymmetric glassy thin-film equation yields a novel, precise, and simple measurement of the surface mobility of glasses. The value found for the latter is compared to values in the literature, and discussed in terms of polymer entanglements and anchoring effects in confinement. Finally, the model predicts a dewetting scenario for ultrathin polymer films, which might have important practical consequences.

\begin{figure*} [th]
\centering
\includegraphics[width=0.5\textwidth]{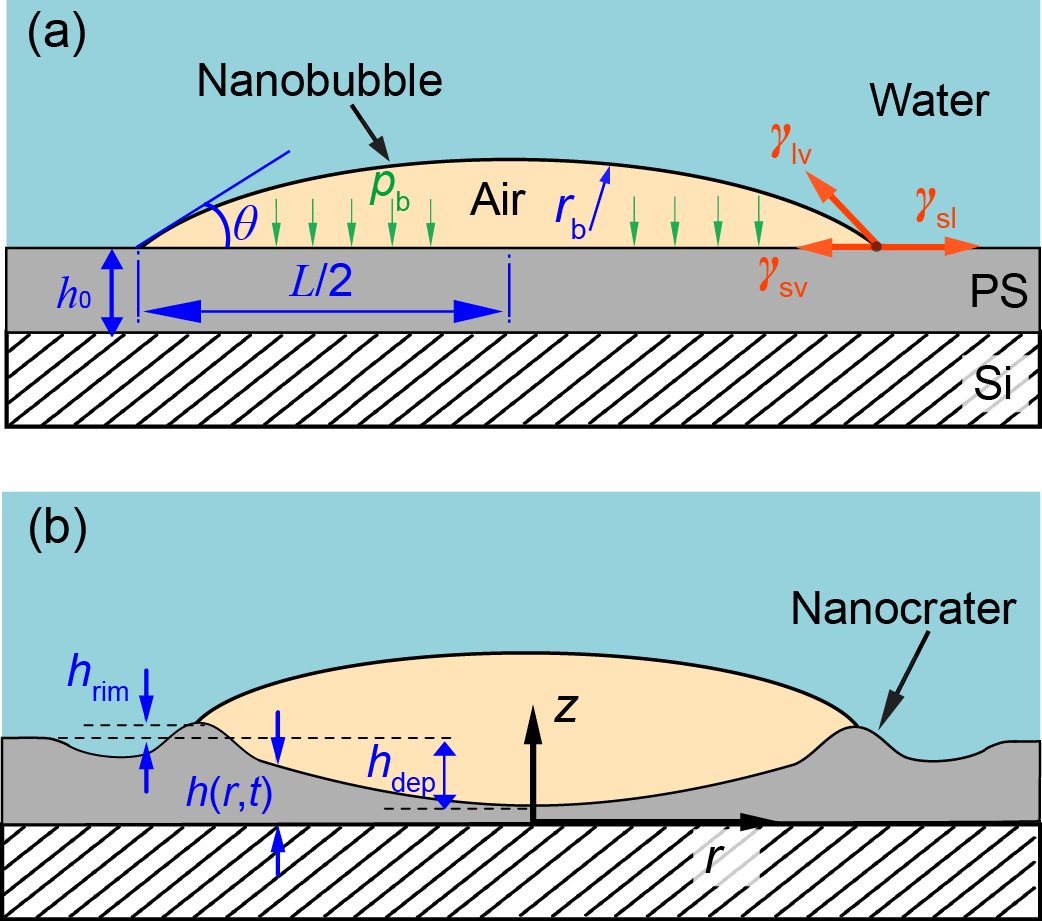}
\caption{Schematic of the system. (a) An air nanobubble spontaneously forms at the PS-water interface right after immersion of the glassy PS sample in water. (b) Subsequently, a nanocrater appears beneath the nanobubble, and grows with time, as the liquid-like layers at both PS-fluid interfaces flow due to the capillary pressure gradient.}
\label{demo}
\end{figure*}

A schematic representation of the system is shown in Fig.~\ref{demo}, where we define the bubble's contact diameter $L$, the bubble's radius of curvature $r_{\textrm{b}}$, the equilibrium contact angle $\theta$, and the initial PS film thickness $h_0$. Note that $L$, $r_{\textrm{b}}$, and $\theta$ are related through volume conservation. According to the Young-Laplace equation, the pressure inside the bubble reads $p_{\textrm{b}} = p_{\textrm{am}} + 2\gamma_{\textrm{lv}}/r_{\textrm{b}}$, where $\gamma_{\textrm{lv}}$ is the water-air surface tension, and $p_{\textrm{am}}$ is the ambient water pressure. In the following, we will quantify how the capillary pressure gradient can lead to the deformation of the glassy PS film, and to the spatiotemporal evolution of the PS nanocrater. The latter is characterized by its depth $h_{\textrm{dep}}$ and rim height $h_{\textrm{rim}}$.

\begin{figure*}[bhtp]
\centering
\includegraphics[width=0.8\textwidth]{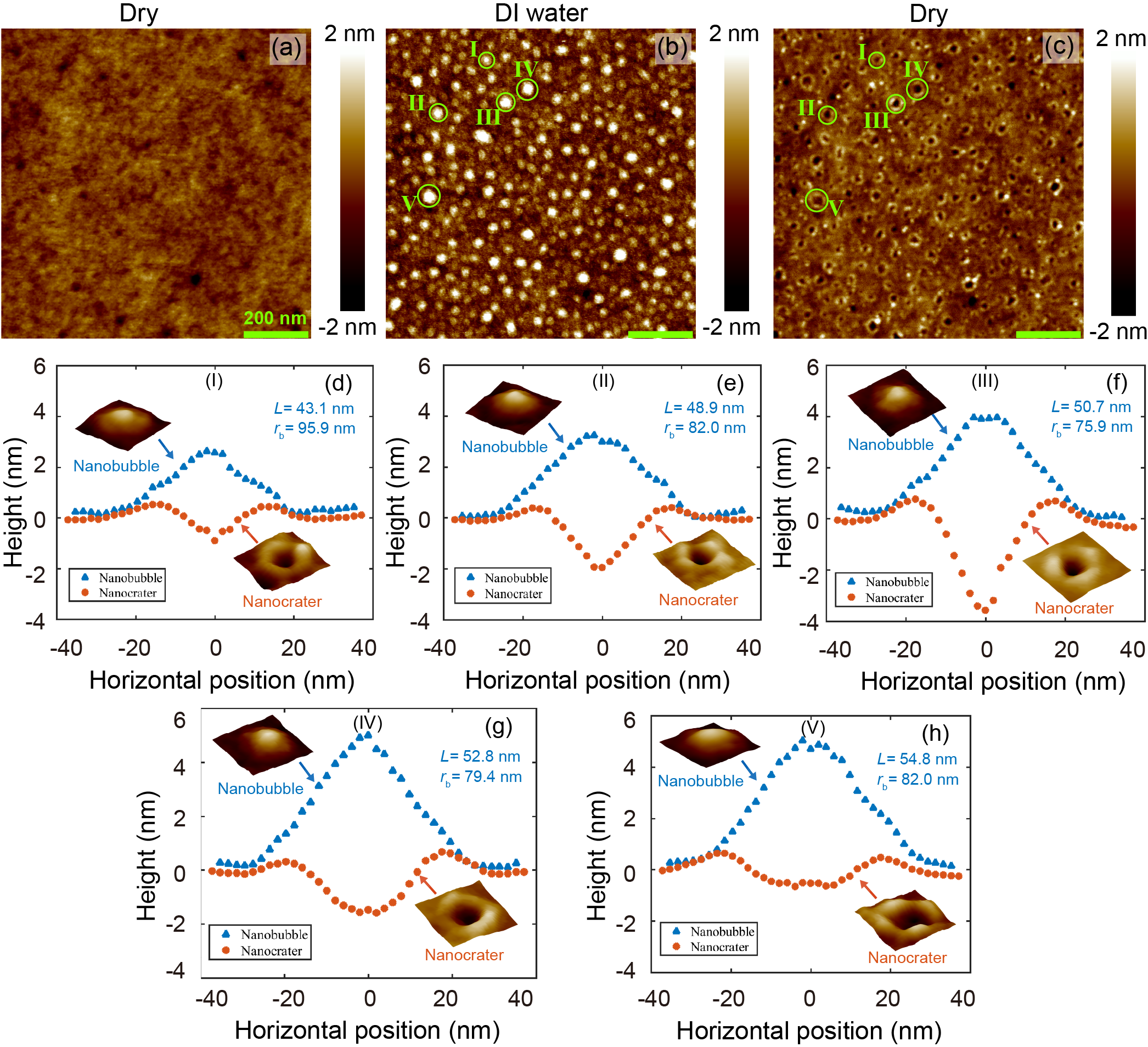}
\caption{Typical AFM images of an ultrathin glassy PS film with thickness $h_0=4.9\pm0.6$~nm in various situations: (a) before immersion in water; (b) after immersion in water, where nanobubbles (white) with an average contact diameter of 50~nm appear on top of it; (c) after immersion in water for $t_{\textrm{b}}\approx240$~min, and subsequent removal of water followed by drying in air during $t-t_{\textrm{b}}\approx$ 250~min. (d-h) Five cross-sectional profiles of nanobubble-nanocrater couples (sorted by increasing $L$ values): $L=$43.1~nm and $r_{\textrm{b}}=95.9$~nm (d, couple I); $L=$48.9~nm and $r_{\textrm{b}}=82.0$~nm (e, couple II); $L=$50.7~nm and $r_{\textrm{b}}=75.9$~nm (f, couple III); $L=$52.8~nm and $r_{\textrm{b}}=79.4$~nm (g, couple IV); $L=$54.8~nm and $r_{\textrm{b}}=82.0$~nm) (h, couple V). The insets in each of those five panels are the 3D AFM images of the nanobubbles and the corresponding nanocraters.}
\label{AFMImg}
\end{figure*}

Ultrathin PS films with three different thicknesses $h_0\in \{2.8\pm0.6, 4.9\pm 0.6, 7.1\pm0.8\}$~nm were prepared by spincoating a solution of PS (Sigma-Aldrich) in toluene onto a silicon wafer, at different toluene mass fractions $\{0.07, 0.10, 0.08\}\,\textrm{wt}\%$ , and with rotational speeds of $\{1200, 1200, 1000\}$~rpm, respectively. The molecular weight of PS is about 350~kg/mol. After spincoating, the PS films were baked inside an oven at a temperature of $45^{\circ}$C for 4~h, in order to evaporate the remaining toluene, and before measuring the thicknesses of the PS films in air, at room temperature, with an AFM (Resolve, Bruker, USA) in tapping mode and a scratching method~\cite{Wang2019}. Figure \ref{AFMImg}a shows a typical AFM image of the PS film in air with a thickness $h_0=4.9\pm 0.6$~nm. The root-mean-squared roughness is about 0.22~nm.

After immersion in deionized (DI) water at room temperature, nanobubbles with diameters ranging from 30~to~100~nm spontaneously nucleated (Fig.~\ref{AFMImg}b) at the PS-water interface~\cite{Wang2017}. The PS sample was kept in water for $t_{\textrm{b}}\approx240$~min, before the water was removed and the sample surface was dried in air for $t-t_{\textrm{b}}\approx250$~min. The same area of the sample was then scanned again with the AFM, as shown in Fig.~\ref{AFMImg}c (see also Fig.~S1 in the Supporting Information (SI)). One observes the existence of nanocraters into the PS film. These nanocraters were generated at the exact same locations where the nanobubbles resided, when the sample was immersed in water.

The cross-sectional profiles for five different nanobubbles and their associated nanocraters (sorted by increasing nanobubble size) are shown in Figs.~\ref{AFMImg}d-h. Interestingly, these profiles qualitatively ressemble the ones obtained on low-molecular-weight PS after embedding and subsequent removal of gold nanoparticles~\cite{Fakhraai2008}. Moreover, it is clear that the lateral sizes of the nanocraters are approximately equal to the sizes of the nanobubbles -- a commensurability valid for all samples in this study (see Fig.~S2 in SI). Nanobubbles with contact diameters $L\le50$~nm typically generate steeper nanocraters, and $h_{\textrm{dep}}$ increases with $L$ for those. (Fig.~\ref{AFMImg}d-f). When the contact diameter $L$ is larger than ~50 nm, the nanocraters are not as curved. Larger bubbles generate shallower craters with decreased $h_{\textrm{dep}}$ and $h_{\textrm{rim}}$ (Fig.~\ref{AFMImg}g-h). With further increased $L$, nanocraters with nearly-flat bottoms are even created (see Fig.~S2i in SI for details).

To rationalize these observations, we invoke a theoretical model that combines two ingredients; i) the existence of a liquid-like layer with viscosity $\eta$ and thickness $h_{\textrm{m}}$ of a few nanometers at the free surface (\textit{i.e.} exposed to any fluid) of the glassy PS film~\cite{Fakhraai2008,Yang2010}; and ii) a lubrication flow in this liquid-like layer~\cite{Chai2014}, driven by the pressure jump between $p_{\textrm{b}}$ and $p_{\textrm{am}}$ at the contact line where the three phases intersect, and opposed by the restoring capillary force due to the induced curvature at the PS-fluid interfaces. Indeed, since the liquid-like layer thickness $h_{\textrm{m}}$ is much smaller than the typical horizontal size $L$, the viscous flow in the layer can be described by lubrication theory~\cite{Batchelor1967}, where the velocity is predominantly in the radial direction, the pressure is constant across the thickness of the liquid-like layer, and the viscous forces therein are balanced by the tangential pressure gradient discussed above. We define $h(r,t)$ as the total thickness profile of the PS film (see Fig.~\ref{demo}), assumed to be axisymmetric given the symmetry of the nanobubble, where $r$ is the horizontal radial spatial coordinate, and $t$ is time. We further assume small slopes for the PS-fluid interfaces, as well as a no-slip boundary condition at the bottom of the mobile layer, located at $z=h(r,t)-h_{\textrm{m}}$, and a no-shear boundary condition at the PS-fluid interfaces, located at $z=h(r,t)$. All together, this leads to the axisymmetric version of the glassy thin-film equation~\cite{Chai2014}, with a novel source term due to the presence of the nanobubble:
\begin{equation}
\frac{\partial h(r,t)}{\partial t} + \frac{h_{\textrm{m}}^{\,3}}{3\eta r} \frac{\partial}{\partial r}\left\{r\frac{\partial}{\partial r}\left[\frac{\gamma_i(r)}{r}\frac{\partial}{\partial r}\left(r\frac{\partial}{\partial r} h(r,t)\right) - p_i(r) \right] \right\} = 0\ ,
\label{eq:gtfe}
\end{equation}
where the surface energy $\gamma_i(r)$ indicates $\gamma_{\textrm{SL}}$ (PS-water) for $r\geq L/2$ and $t<t_{\textrm{b}}$, as well as $\gamma_{\textrm{SV}}$ for either $t>t_{\textrm{b}}$, or $t<t_{\textrm{b}}$ and $r < L/2$; while the external pressure $p_i(r)$ indicates $p_{\textrm{b}}$ for $r\leq L/2$ and $t<t_{\textrm{b}}$, as well as $p_{\textrm{am}}$ for either $t>t_{\textrm{b}}$, or $t<t_{\textrm{b}}$ and $r > L/2$. Due to the constant liquid-like layer thickness $h_{\textrm{m}}$, the equation is linear, and formally ressembles the capillary-driven thin-film equation for bulk flow under perturbative profile variations~\cite{Salez2012a,Backholm2014}. Just before the formation of the nanobubble (assumed to be instantaneous), the PS film has a uniform thickness $h(r,t=0)=h_0$, which we use as an initial condition.

We now nondimensionalize Eq.~\eqref{eq:gtfe} by rescaling the variables through $h=H\,h_0$, $r=R\,L/2$, $t=T\,3\eta L^4 / (16\gamma_{\textrm{SV}}h_{\textrm{m}}^{\,3})$, and  $t_{\textrm{b}}=T_{\textrm{b}}\,3\eta L^4 / (16\gamma_{\textrm{SV}}h_{\textrm{m}}^{\,3})$, which leads to the dimensionless form of Eq.~\eqref{eq:gtfe}:
\begin{equation}
\frac{\partial H(R,T)}{\partial T} + \frac{1}{R} \frac{\partial}{\partial R}\left\{R\frac{\partial}{\partial R}\left[\frac{1-\alpha(T)\Theta(R-1)}{R}\frac{\partial}{\partial R}\left(R\frac{\partial}{\partial R} H(R,T)\right) - \beta(T)\Theta(1-R) \right] \right\} = 0\ ,
\label{eq:gtfeadim}
\end{equation}
where $\Theta$ is the Heaviside function, $\alpha (T) = (\gamma_{\textrm{SV}} - \gamma_{\textrm{SL}})\Theta(T_{\textrm{b}}-T) / \gamma_{\textrm{SV}}$ and $\beta (T)= L^2\gamma_{\textrm{LV}}\Theta(T_{\textrm{b}}-T) / (2h_0r_{\textrm{b}}\gamma_{\textrm{SV}})$. We solve Eq.~\eqref{eq:gtfeadim} numerically from the initial condition $H(R,T=0)=1$, by using a finite-element method where the equation is divided into two coupled second-order partial differential equations involving two fields~\cite{Pedersen2019}: the height $H(R,T)$ and the total pressure $P(R,T)=\frac{\alpha(T)\Theta(R-1)-1}{R}\frac{\partial}{\partial R}\left[R\frac{\partial}{\partial R} H(R,T)\right] + \beta(T)\Theta(1-R)$. The fields are discretized with linear elements, and the coupled equations are solved with a Newton solver from the FEniCS library~\cite{Logg2012}. The numerical routine is performed with a constant time step $\Delta T = 5\cdot 10^{-4}$ and a uniform spatial discretization step $\Delta R = 5\cdot 10^{-4}$. Finally, as spatial boundary conditions at large $R$, we impose both fields to reach $1$, while their first spatial derivatives vanish.

\begin{figure}[t]
\centering
\includegraphics[width=1.0\columnwidth]{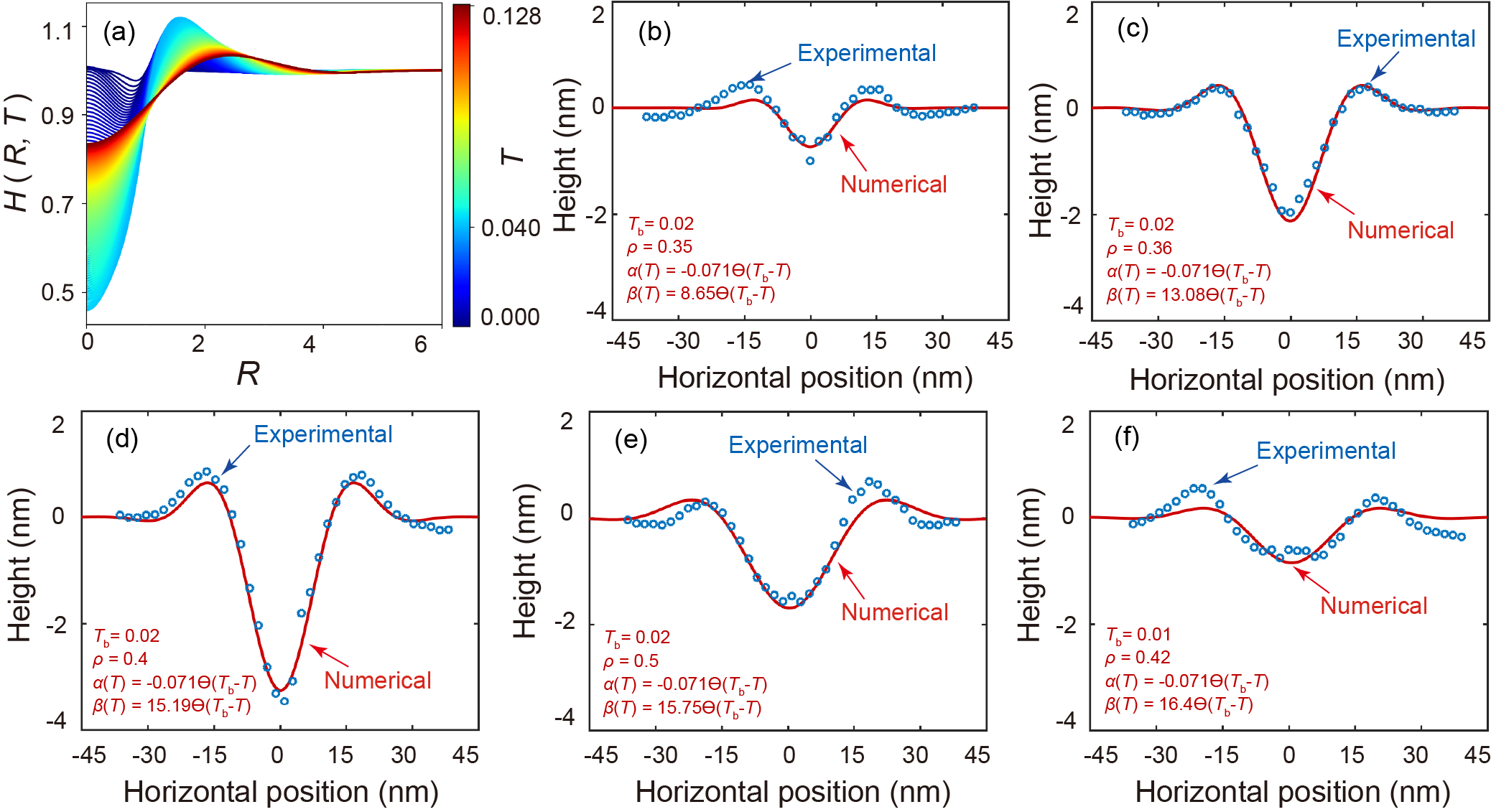}
\caption{Deformation of the PS film: (a) Numerical solution of Eq.~\eqref{eq:gtfeadim} over a time interval $T\in[0,0.128]$, for $T_{\textrm{b}}=0.01$, $\alpha(T)= -0.071 \,\Theta(T_{\textrm{b}}-T)$, and $\beta(T)=5\,\Theta(T_{\textrm{b}}-T)$. (b-f) Cross-sectional AFM profiles (blue circular markers) in air for the five selected nanocraters in Fig.~\ref{AFMImg}d-h, and corresponding best fit (red solid curves) to the numerical solution with $T_{\textrm{b}}=0.02$, $\rho=0.35$, $\alpha(T)= -0.071\, \Theta(T_{\textrm{b}}-T)$, and $\beta(T)=8.65\,\Theta(T_{\textrm{b}}-T)$ (b); $T_{\textrm{b}}=0.02$, $\rho=0.36$, $\alpha(T)= -0.071\, \Theta(T_{\textrm{b}}-T)$, and $\beta(T)=13.08\,\Theta(T_{\textrm{b}}-T)$ (c); $T_{\textrm{b}}=0.02$, $\rho=0.4$, $\alpha(T)= -0.071\, \Theta(T_{\textrm{b}}-T)$, and $\beta(T)=15.19\,\Theta(T_{\textrm{b}}-T)$ (d); $T_{\textrm{b}}=0.02$, $\rho=0.5$, $\alpha(T)= -0.071\, \Theta(T_{\textrm{b}}-T)$, and $\beta(T)=15.75\,\Theta(T_{\textrm{b}}-T)$ (e); $T_{\textrm{b}}=0.01$, $\rho=0.42$, $\alpha(T)= -0.071\, \Theta(T_{\textrm{b}}-T)$, and $\beta(T)=16.4\,\Theta(T_{\textrm{b}}-T)$ (f). Note that the horizontal and vertical origins are arbitrarily shifted.}
\label{fig:growth rates}
\end{figure}

Figure~\ref{fig:growth rates}a shows an example of a numerical solution of Eq.~\eqref{eq:gtfeadim}. One can see that both the dimensionless depth $h_{\textrm{dep}}/h_0$ of the nanocrater and the dimensionless height $h_{\textrm{rim}}/h_0$ of the rim increase with dimensionless time. As the fluid in the liquid-like layer gets displaced, we also observe a continuous lateral shift in the dimensionless horizontal position of the rim.

In Fig.~\ref{fig:growth rates}b-f, we fit the numerical solutions to the experimental profiles, for five nanocraters created by the five selected nanobubbles (shown in Fig.~\ref{AFMImg}d-h) of increasing contact diameters $L$ from b to f. To do so, we first put back dimensions in the numerical solutions, by using the experimental parameters: $t-t_{\textrm{b}}=250$~min, $\gamma_{\textrm{SV}}=40.7$~mN/m, $\gamma_{\textrm{LV}}=$72.8~mN/m, and $h_0=4.9\pm0.6$~nm, as well as the values of $t_{\textrm{b}}$, $L$ and $r_{\textrm{b}}$ for each bubble. In order to account for geometrical uncertainties, we also include a prefactor $\rho$ in both $L$ and $r_{\textrm{b}}$ as a fitting parameter, which for all experiments in this study is found in the range $0.3\leq \rho \leq 0.5$.
We observe that the numerical solutions show a good agreement with experimental cross-sectional profiles for all five exemplary nanocraters. The depth $h_{\textrm{dep}}$ of the nanocraters first increases and then decreases with increasing $L$. Interestingly, we find that it is actually $r_{\textrm{b}}$ that determines $h_{\textrm{dep}}$. With increasing $L$, $r_{\textrm{b}}$ first decreases from 95.9 nm (bubble I) to 75.9 nm (bubble III). Then it increases from 75.9 nm (bubble III) to 82.0 nm (bubble V). The smaller $r_{\textrm{b}}$ leads to the larger deformation in the PS film, \textit{i.e.} the larger magnitudes of the rim height $h_{\textrm{rim}}$ and crater depth $h_{\textrm{dep}}$. This is expected due to the Laplace pressure of the nanobubbles, that scales as $\sim 1/r_{\textrm{b}}$, and that drives the deformation of the PS layer.

From the fitting procedure detailed above, we extract a single relevant free parameter: the surface mobility $h_{\textrm{m}}^3/(3\eta)= 2.31^{+1.73}_{-1.92} \times 10^{-10}~\textrm{nm}^3/(\textrm{Pa.s})$ of 350~kg/mol PS. Regardless of the total PS film thickness, and the nanobubble geometry, the different experiments self-consistently exhibit the same value of surface mobility. Previous measurements on low-molecular-weight PS, at higher temperatures, exhibited an Arrhenius-like trend for the temperature dependence of the surface mobility, which is characteristic of a liquid-like behaviour~\cite{Chai2014}. Interestingly, the extrapolation of the latter empirical behaviour to room temperature would lead to a surface mobility ten times lower than the one reported here, despite the much higher molecular weight used here which would have suggested the opposite due to reptation and anchoring effects~\cite{Chai2019}. This brings two possible non-exclusive scenarios: i) a saturation of the surface mobility at low temperature~\cite{Fakhraai2008}; ii) a reduction of the entanglement density~\cite{Silberberg1982,Si2005,Baumchen2009}, and thus viscosity~\cite{Brochard2000,Bodiguel2006,Shin2007}, in strong confinement.

Finally, we want to stress that the PS deformation profiles are transient, and that they in fact will continue to evolve with increasing time (see Fig.~\ref{fig:growth rates}a), although very slowly. Moreover, a careful mathematical analysis of Eq.~\eqref{eq:gtfe} reveals the absence of any relevant stationary state, which implies a dramatic consequence: due to the existence of a liquid-like surface layer, and provided the films are thin enough (\textit{i.e.} $h_0$ close to $h_{\textrm{m}}$) to avoid anchoring effects at large molecular weights~\cite{Chai2019}, the presence of surface nanobubbles should eventually lead to the dewetting of any ultrathin glassy PS film~\cite{Reiter2001,Damman2003}. The critical time for dewetting is solely controlled by the parameters $\theta$, $\gamma_{\textrm{SV}}$, $\gamma_{\textrm{LV}}$, $h_0$, and $L$ (or $r_{\textrm{b}}$, due to volume conservation) above, as well as the surface mobility $h_{\textrm{m}}^3/(3\eta)$.

As a conclusion, we have shown that immersing ultrathin glassy polystyrene films in water, in ambient conditions, leads to the spontaneous nucleation of air nanobubbles, which then generate nanocraters into the free surface of the PS films. The mechanism of such a dynamical deformation process is found by combining experimental atomic-force microscopy with a mathematical model based on lubrication theory applied to the liquid-like layer present at the free surface of a glassy film. The liquid-like layer is driven to flow by the pressure jump at the contact line where the three phases intersect, between the nanobubble's inner Laplace pressure and the outer ambient pressure, and opposed by the capillary force due to the induced curvature at the PS-fluid interfaces. Since the Laplace pressure scales as the inverse of the bubble's radius of curvature, the size of the nanocraters can be finely controlled. From the excellent agreement between the experimental profiles and the numerical solutions of the modified glassy thin-film equation, we extract the surface mobility of the glassy films. Comparison of the surface mobility with extrapolated results from the literature points towards the possible saturation of surface mobility at low temperature, and/or the reduction of polymeric entanglement density (and thus viscosity) in confinement. Our work thus provides a novel, precise, and simple measurement of the surface nanorheology of glasses. Furthermore, our results highlight the influence of surface nanobubbles on the stability of immersed ultrathin glassy polymer films: the nanobubbles can drive the film towards dewetting, which would have important consequences for nanoimprint lithography~\cite{Teisseire2011} and nanomechanical data storage~\cite{Vettiger2002}, to name a few.
\newline

\begin{acknowledgements}
The authors thank James Forrest, Robert Style, Detlef Lohse and Jacco Snoeijer for interesting discussions. Y. W. and S. R. appreciate financial support from the National Natural Science Foundation of China (Grant No. 51775028) and the Beijing Natural Science Foundation (Grant No. 3182022).
\end{acknowledgements}
\bibliographystyle{apsrev4-1}
\bibliography{Ren2020}
\end{document}